\documentclass[12pt]{naturenew}
\usepackage{ulem}
\usepackage{amsmath}
\usepackage{color}
\usepackage{graphicx}
\usepackage{caption}
\usepackage{subcaption}
\usepackage{epsfig}
\usepackage{SIunits}
\usepackage{braket}
\newlength{\upit}\upit=0.1truein

\newcommand{\ltappr}{{{\lower4pt\hbox{$<$} } \atop \widetilde{ \ \ \ }}}
\newlength{\bxwidth}\bxwidth=1.5 truein

\newcommand{\dg}{^{\dagger}}

\newcommand{\be}{\begin{equation}}
\newcommand{\ee}{\end{equation}}
\newcommand{\ba}{\begin{eqnarray} }
\newcommand{\ea}{\end{eqnarray}}

\newlength{\figwidth}
\newlength{\shift}
\newlength{\fight}

\newcommand{\bk}{{\bf{k}}}

\newsavebox{\fmbox}

\newcommand{\spm}{s$^{\pm}$}

\newcommand{\zx}{\hbox{zx}}
\newcommand{\zy}{\hbox{zy}}

\newcommand{\isop}{{ I}}

\title{Entangled Orbital Triplet Pairs in Iron-Based Superconductors}

\author
{T. Tzen Ong $^{1}$, P. Coleman $^{1,2}$ and J. Schmalian $^{3,4}$}
\begin{document}
\maketitle
\begin{affiliations}
\item Center for Materials Theory, Department of Physics and Astronomy,
Rutgers University, 136 Frelinghuysen Rd., Piscataway, NJ 08854-8019, USA
\item Department of Physics,  Royal Holloway University
of London, Egham, Surrey TW20 0EX, UK.
\item Institute for Theory of Condensed Matter, KIT,
76131 Karlsruhe, Germany.
\item DFG Center for Functional Nanostructures,
KIT, 76128 Karlsruhe, Germany.
\end{affiliations}

\begin{abstract}
A key question in high temperature iron-based
superconductivity is the mechanism by which the paired electrons
minimize their strong mutual Coulomb repulsion.  While
electronically paired superconductors generally avoid the Coulomb
interaction through the formation of nodal, higher angular momentum pairs,
iron based superconductors appear to form  singlet
s-wave (s$^{\pm}$) pairs.
By taking the orbital degrees of freedom of the iron atoms into account,
here we argue that the s$^{\pm}$ state in
these materials possesses internal d-wave structure,
in which a  relative  d-wave ($L=2$) motion of the pairs entangles with
the ($I=2$) internal angular momenta of the d-orbitals
to form a low spin $J=L+I=0$ singlet.
We discuss how the recent observation of a nodal gap with
octahedral structure in
KFe$_{2}$As$_{2}$\cite{ShinScience2012,TailleferNatPhys2013}
can be understood as a high spin ($J=L+I=4$) configuration of the
orbital and isospin angular momenta;
the observed
pressure-induced phase transition into a fully gapped
state\cite{TailleferNatPhys2013} can then interpreted as a high-to-low
spin phase transition  of  the Cooper pairs.
\end{abstract}


The family of iron-based high temperature superconductors exhibits a
marked absence of nodes in the pair wavefunction \cite{EisakiTerasakiJPSJ09,EisakiYashimaJPSJ09,MatsudaPRL2009,
MolerJPSJ09,mazinschmalian2009,TakeuchiPRL09,TsueiNatPhys10,HanaguriScience10,
MazinRepProgPhys11,KAMPRL2011,HoffmanRPP11,AllanScience12,HongEPL12}.
This stands in stark contrast with almost all other strongly correlated
superconductors and superfluids, including the cuprates, heavy
fermions, ruthenates and $^{3}$He \cite{Jerome2008,BatailPRL00,PfleidererRMP09,Tsuei2000,Mack03,RiceSigrist1995,MaenoNature1998,
SigristNature1998, MacKenzieMaenoRMP2003,BalianWerthamerPhysRev1963,AndersonPhysRev1961,
OsheroffRevModPhys1997,LeggettRMP1975,WheatleyRMP1975,VollhardtWolfle2013},
where the repulsive interaction between fermions drives the formation of higher angular momentum pairs with nodes in the pair wavefunction. In distinction, the Fe-based superconductors   are generally believed to have an isotropic $s^{\pm}$ structure.
The underlying concern, voiced by Lev D. Landau[see: V. L. Ginzburg, {\em About Science, Myself and Others } CRC Press (2004)], that one cannot repeal Coulomb's law must therefore find a new resolution in the Fe-based systems and other unconventional superconductors with multiple orbitals.
In this context it is interesting that
recent experiments\cite{ShinScience2012,TailleferNatPhys2013,
DrechslerPRB2013,WatanabePRB2014,ShinPRL2012,DingPRB2013,ShinPRB2014,MachidaJPSJ2014}
show that upon doping, the gap structure can undergo a
sudden transformation into an anisotropic paired state, suggesting the formation of higher angular
momentum pairs.  Is this a consequence of a competition between s-wave
and higher angular momentum channels, or can a single unifying pairing
mechanism account for these disparate experimental results?  Here we
show the existence of such a mechanism.  By taking into account the
unique helical orbital structure of the electronic bands,
we show that an underlying $d$-wave orbital
triplet can give rise to a new type of $s^{\pm}$ state.  Furthermore,
we show that transformations in the relative orientation of the
orbital and atomic isospin angular momenta of the pairs can account
for the observed transition from an isotropic $s^{\pm}$-wave to an
octet gap structure. An implication of our mechanism is that the origin or the pairing state is in the formation of hidden $d$-wave pairing.

The key to our theory lies in the
helical orbital structure of the electronic bands, in which the
orbital character of the quasiparticles behaves as a vector in orbital
space, rotating twice as one passes
around the $\Gamma$ point in the Brillouin zone\cite{DHLeePRB2009},
as shown in Fig. ~\ref{Fig:Orbital isospin}.
This topologically non-trivial  band structure
is well established from first-principles calculations
\cite{KotliarHauleYinNMat2011} and has been independently confirmed
by ARPES measurements in both
the normal state\cite{FengZhangPRB2011,HasanXiaPRL2009} and the
spin-density wave phase\cite{DingPRL2010, HsieharXiv0812.2289}.
The dominant atomic orbital character of
Bloch waves near the Fermi surface involves the
three $t_{2g}$ orbitals, i.e. the $xz$, $yz$ and
$xy$ orbitals\cite{LeeWenPRB2008}.
To illustrate the key elements of our theory,
we adopt a simplified two orbital ($xz/yz$) model which captures the
orbital helicity of the bands \cite{Raghu:2008ta,DHLeePRB2009};
later inclusion of the $xy$ orbitals does not change the key conclusions.
The $xz$ and $yz$  orbitals form a degenerate doublet
or ``iso-spin'' which we label with the $\isop_{z}$ index, treating the
$xz$ orbital as an ``up'' state with $\isop_{z}=+1$ and the $yz$
orbital as a ``down'' state with $\isop_{z}= - 1$.
The electrons in these orbitals  carry
internal $L = 2$ angular momentum, containing a mixture of
$M_{z}=\pm 1$ states.
There are thus two potential sources of angular momentum
carried by the Cooper pairs:
external ($\hat L$) ``orbital'' angular momentum associated with the
relative electron motion and
internal ``isospin'' angular momentum ($\hat I$) associated with the
atomic electron states.

As electrons that form Cooper pairs hop between sites on the lattice,
they exchange $\pm $ 2$\hbar $ units of angular momentum between
the orbital and isopin angular momentum. These ``isospin-flip''
hopping processes are the analogue of the spin-orbit coupling terms
in metals which give rise to Rashba coupling terms. We write down
the tight-binding two-orbital Hamiltonian $H_{0} (\bk )$
for the electron motion\cite{Raghu:2008ta}
in a fashion that highlights the isospin-orbital
Rashba coupling,
\begin{eqnarray}
\label{eq:Orbital Rashba}
\hat H_0 (\bk ) &=&
 \epsilon_{s} (\bk ) \mathbf{1} - \vec B_{\bk }\cdot \vec{ \isop }.
\end{eqnarray}
Here we have introduced a triplet of
isospin Pauli matrices $\vec{\isop}= (\isop_{1},\isop_{2},\isop_{3})$
which span the orbital space, defined as
$\isop_{1}\equiv \ket{xz} \bra{yz} + \ket{yz} \bra{xz} $,
$\isop_{2}\equiv -i\ket{xz} \bra{yz} + i\ket{yz} \bra{xz} $
and $\isop_{3}\equiv \vert \zx\rangle\langle \zx\vert- \vert
\zy\rangle\langle \zy\vert$. The orbital Rashba field
$\vec{B}_{\bk } = (\epsilon_{xy} (\bk ), 0, \epsilon_{x^{2}-y^{2}}
(\bk ))= B_{\bk } \, \hat{\bf{ n } }_{\bk}$
has magnitude
$B_{\bk }=\sqrt{\epsilon_{x^{2}-y^{2}} (\bk)^{2} + \epsilon_{xy} (\bk )^{2}}$
and direction
$\hat{\bf n}_{\bk }=  (\sin \phi_{\bk },0,\cos \phi_{\bk })$
where $\phi_{\bk }= \tan^{-1}(\epsilon_{xy} (\bk)/\epsilon_{x^{2}-y^{2}} (\bk ))$
is the clockwise angle of rotation about the $\hat y$ axis. The transformation behavior of the first and third components of $\hat{\bf n}_{\bk }$ is dictated by the point group transformation, while the vanishing second component of $\hat{\bf n}_{\bk }$ is a consequence of time reversal symmetry.
Electron hopping between iron atoms proceeds
predominantly via the arsenic atoms, resulting
in a preferential hopping of $yz$-orbitals along the $x$ axis and
$xz$-orbitals along the $y$ axis.
The corresponding tight-binding description\cite{Raghu:2008ta}
then gives $\epsilon_{s} (\bk )= 4 t_{2} (c_{x} c_{y}) + 2t_{1} (c_{x}+c_{y})$,
$\epsilon_{xy} (\bk ) = 4t_{4} s_{x}s_{y}$ and $\epsilon_{x^{2}-y^{2}} (\bk )= 2t_{3} (c_{x}-c_{y})$.

It is straightforward to diagonalize the Hamiltonian to obtain
two quasi-particle bands of opposite helicities, $\isop = \pm 1$,
where $\isop$ is the eigenvalue of the
``helicity'' operator $\vec{\isop}\cdot \hat{\bf {n}}_{\bk }$.
The energies are given by
$E_{\isop}(\bk) = \epsilon_{s} (\bk )- {\rm sgn} (\isop) B_{\bk }$,
giving rise to a normal state Hamiltonian
\begin{equation}
\label{qp hamiltonian}
H_{0} = \sum_{\bk  , \sigma } (E_{+}(\bk) a\dg_{\bk \sigma }a_{\bk \sigma } + E_{-}(\bk) b\dg_{\bk \sigma }b_{\bk \sigma })
\end{equation}
where
\begin{eqnarray}\label{l}
a\dg _{\bk , \sigma}&=& u_{\bk }c\dg _{\bk, xz, \sigma } + v_{\bk
}c\dg _{\bk, yz, \sigma }, \cr
b\dg _{\bk , \sigma} &=& u_{\bk }c\dg
_{\bk, yz, \sigma }- v_{\bk } c\dg _{\bk, xz, \sigma }
\end{eqnarray}
are the hole and electron quasi-particle creation operators respectively.
The helicity has an ``\spm'' symmetry,
with $\isop = + 1$ on the hole pockets and $\isop = - 1$ on
the electron pockets.  The quasi-particle coherence factors
$(u_{\bk},v_{\bk })= (\cos \phi_{\bk}/2, \sin \phi_{\bk }/2)$
are determined by the orientation of the $\hat {\bf n}_{\bk }$ vector,
which winds twice in isospin space
as one passes around the  $\Gamma$ point (see Fig. ~\ref{Fig:Orbital isospin}).
The vector $\vec{n}_{\bk}$ reverses its direction of
rotation around the $(\pi, \pi)$ point, and when this hole pocket is
translated into the central zone, it forms a second $\Gamma$ pocket
with an opposite orbital character to the first.




The multi-orbital nature of the band structure
suggests that the gap function is entangled with the orbital
isospin. To this end we make the ansatz that
the \spm gap alternates between the electron and hole pocket,
but is  constant on each of these Fermi surfaces, given by
a pairing Hamiltonian
\begin{equation}\label{}
H_{sc} = \Delta \sum_{\bk  }\left[
a\dg_{\bk \uparrow}a\dg_{-\bk \downarrow}
- b\dg_{\bk \downarrow} b\dg_{-\bk  \uparrow}+ \rm {H.c} \right]
\label{spmm},
\end{equation}
where for simplicity, we have chosen the two gaps to be of equal
magnitude.
In a conventional picture of \spm\ pairing, the gap function
$\Delta^{\pm}(\bk) \sim  \Delta_0 \cos(k_x) \cos(k_y)$
on all bands.
As we now show, in this alternative interpretation of \spm\ pairing, a condensate of orbitally
entangled d-wave pairs hides behind the topologically non-trivial band structure.

The orbital Rashba field $\hat n_{\bk }$  shown in Fig.~\ref{Fig:Orbital isospin},
defines a quasiparticle reference frame which rotates in orbital space
as it moves through momentum space.
Though the \spm\ gap is constant in the quasiparticle reference frame,
when transformed back into the stationary orbital reference frame,
a non-trivial orbital and isospin structure is revealed.
Using equations (3), carrying out the
transformation back into the fixed atomic orbital basis, we obtain
\begin{eqnarray}
\label{eq:TAO pairing}
H_{sc} & = & -\Delta \sum_{\bk }c\dg_{\bk\uparrow}
\left[(u_{\bk }^{2}-v_{\bk }^{2})
\isop_{3}
 + 2 u_{\bk }v_{\bk } \isop_{1}
 \right]c\dg_{-\bk \downarrow}\cr
       & = & \frac{1}{2}\sum_{\bk } c\dg_{\bk}
\left[(\Delta _{x^{2}-y^{2}} (\bk )\isop_{3}
+ \Delta_{xy} (\bk )\isop_{1}
 \right] i\sigma_{2}c\dg_{-\bk}
\end{eqnarray}
where we have used the fact that $u_{\bk }$ and $v_{\bk }$ are odd
functions  of momentum.
Here
$\Delta_{xy} (\bk ) = -\tfrac{\Delta }{|B_{\bk }|}\epsilon_{xy} (\bk)$ and
$\Delta_{x^{2}-y^{2}} (\bk ) =-\tfrac{\Delta
}{|B_{\bk}|}\epsilon_{x^{2}-y^{2}} (\bk)$ define d-wave form-factors.
In this way, the gap function is revealed to be a triplet
pair wavefunction, reminiscent of superfluid He-3B,
except that it involves isospin rather than spin
operators, and the
gap functions have d-wave rather than p-wave
form-factors. This leads us to identify  the \spm\ gap as
a ``d-wave orbital triplet''\cite{OngColemanPRL2013} condensate.

To further elucidate the orbital triplet condensate,
we combine $H_{0}+H_{sc}$ to obtain
\begin{equation}
\label{eq:TAO Hamiltonian}
H = \frac{1}{2}\sum_{\bk }\psi\dg _{\bk }[(\epsilon_{s} (\bk )
- \vec{B}_{\bk }\cdot \vec{I})\gamma_{3}
+ \vec{d}_{\bk }\cdot \vec{\isop}\gamma_{1}
]
\psi_\bk,
\end{equation}
where we have used a four-component Balian Werthammer\cite{BalianWerthamerPhysRev1963}
notation, $\psi\dg _{\bk } =
\left( c\dg _{\bk \isop \sigma},
 c_{-\bk \isop \sigma' } (i\sigma_{2})_{\sigma' \sigma}\right)$,
using $\isop$ and $\sigma$ denote the orbital and spin quantum numbers
and $\vec{\gamma}= (\gamma_{1},\gamma_{2},\gamma_{3})$ for the $2 \times 2$ Nambu  matrices acting in particle-hole space.
The Bogoliubov spectrum is then given by,
\be
\label{app eq: Bogoliubov bands}
E^{\alpha}(\bk) = \left[ (\epsilon_{\bk}^2 + |\vec{B}_{\bk}|^2 + |\vec{d}_{\bk}|^2) - \alpha \sqrt{ 4 \epsilon_{\bk}^2 |\vec{B}_{\bk}|^2 + 4 |\vec{B}_{\bk} \times \vec{d}_{\bk}|^2} \right]^{1/2}
\ee
where, in analogy to $^{3}$He-B, we have defined a $\vec{d}$-vector
for the orbital triplet pairing,
\ba
\label{eq:d vector}
\vec{d}_{\bk }  =  (\Delta_{xy} ,\  0,\  \Delta_{x^2-y^2}),
\ea
Like the orbital Rashba vector $\vec{B}_{\bk }$, the $\vec{d}$-vector
precesses in the x-z plane of isospin space, its
d-wave form factor guaranteeing that it rotates twice  as $\bk $ goes
around the $\Gamma$ point.
The vanishing second component of $\vec{d}$ is a consequence of time reversal symmetry of the pairing state.
As long as there is no relative motion between $\vec{B}_{\bk }$ and
$\vec{d}_{\bk }$, the gap function preserves its phase and the
underlying  nodes of the $d$-wave form factor are ``hidden".
This superconductor, in which the two d-wave condensates are locked in phase,
is normally favored by its large isotropic gap, and corresponds to a
low spin  ($J = L - I = 2-2 = 0$) $s^{\pm}$ condensate.
We note as an aside that were $^{3}$He-B to contain an analogous
spin Rashba term, its Fermi surface would also split into two
components with an \spm\  structure \cite{OngColemanarXiv:1402.7372}.

Our picture allows us to consider generalizations in which the
relative sign of the two d-wave components is reversed.
We will show that in the case where the electron or hole Fermi pockets are
uncompensated, the Coulomb interaction changes the sign of the
Josephson coupling between the two condensates, driving this reversal.
When the two condensates have opposite phase, $\vec{d}_{\bk}$
rotates oppositely  to $\vec{n}_{\bk}$, and since each vector counter rotates
twice passing around the $\Gamma$ point, in the quasiparticle
reference frame the relative phase between the two
vectors rotates four times as one passes around the $\Gamma$ point,
giving rise to a ``high spin'' gap function with g-wave $J= L + I = 4$
total angular momentum and an octet gap structure.
To reveal the octet gap structure, we
reverse the sign of $\Delta_{x^2-y^2}$ in Eq.~\ref{eq:TAO pairing} and
transforming back to the quasiparticle basis, to  obtain
\begin{eqnarray}
H_1 & = & \sum_{\bk} c^{\dagger}_{\bk} [
\Delta_{xy} \isop_1-\Delta_{x^2-y^2} \isop_3  ] (i\sigma_2) c\dg _{-\bk} + {\rm H.C.} \\ \nonumber
    & = &- \frac{1}{B_{\bk}} \left( \Delta_{xy} \epsilon_{xy} - \Delta_{x^2-y^2} \epsilon_{x^2-y^2} \right) \left[ a^{\dg}_{\bk \uparrow} a^{\dg}_{-\bk  \downarrow} - b^{\dg}_{\bk  \uparrow} b^{\dg}_{-\bk  \downarrow} \right] \\ \nonumber
    &   & -\frac{1}{B_{\bk}} \left( \Delta_{xy} \epsilon_{x^2-y^2} + \Delta_{x^2-y^2} \epsilon_{xy} \right) \left[ a^{\dg}_{\bk  \uparrow} b^{\dg}_{-\bk  \downarrow} - b^{\dg}_{\bk  \uparrow} a^{\dg}_{-\bk  \downarrow} \right]+{\rm H.C}
    \label{eq:Octet SC}
\end{eqnarray}
The octet gap  symmetry is revealed by setting
$\Delta_{xy} (\bk ) \sim  \epsilon_{xy}$ and
$\Delta_{x^{2}-y^{2}} (\bk )  \sim
\epsilon_{x^{2}-y^{2}} $,
for which the band-diagonal component of the pairing
is given by $\Delta (\bk ) \propto
\left(\epsilon_{xy}^2 -\epsilon_{x^{2}-y^{2}}^2
\right)\sim  \cos(4\theta)$, corresponding
to a gap
with eight nodes
whose exact positions are determined by the ratio of gap magnitudes $\Delta_{2}/\Delta_{1}$.
There is also an
inter-band A$_{2g}$ pairing term, given by
$\vec{B}_{\bk} \times \vec{d}_{\bk} \propto \left(
\Delta_{xy} \epsilon_{x^2-y^2} - \Delta_{x^2-y^2} \epsilon_{xy}
\right)$
which produces a small
second-order correction to the gap which does not alter its basic symmetry.

The internal $d$-wave structure of the orbital triplet
will always act to minimize the total Coulomb repulsion;
however, the orbital Rashba terms will in general mix
the $d$-wave orbital triplet pairing with conventional $s^{++}$ pairing.
In electron-hole pocket materials,
the phase cancellation of the electron and hole pockets
(and opposite helicities) minimizes the on-site $s$-wave component
induced by the orbital Rashba terms in the kinetic energy, thereby minimizing the
on-site Coulomb repulsion at iron sites. However,
in systems with just electron or hole pockets,
this cancellation fails. In this situation we expect the
orbital triplets to accommodate the Coulomb interaction by
reversing the helicity of the $\vec{d}$ vector, building explicit
octet nodes into the gap function.

Fig.~\ref{Fig:Octet SC gap} shows the octet superconducting gap around
the hole pockets around $\Gamma$ for the high
angular momentum case. Recent ARPES
experiments on KFe$_2$As$_2$
show evidence for the high angular
momentum octet superconducting state on the hole pockets\cite{ShinScience2012}, which is
confirmed to have nodes via thermal conductivity
measurements\cite{TailleferNatPhys2013}. Our orbital triplet scenario
is consistent with these observations.
Fig.~\ref{Fig:Octet SC gap} shows the gap in the vicinity of
the electron pockets around $M$.
Orbital triplet pairing also reproduces the large isotropic gap
seen experimentally in strongly  electron doped materials, where
generically,  the octet line nodes do not intersect the electron
pockets, leading to a full gap.

We now discuss the nature of the phase transition between the low-spin
\spm and the high spin octet state.
The basic structure of the phase transition can be modeled
using a Landau Free energy,  which we write as $F=F_{DOT}+F_{S}$, where
\ba
\label{eq:GL}
F_{DOT} & = & \alpha (T-T_{c}) (|\Delta_1|^2 + |\Delta_2|^2) - \chi_{12} \Delta_1 \Delta_2 + \beta_1 (|\Delta_1|^4 + |\Delta_2|^4) + \beta_2 |\Delta_1|^2 |\Delta_2|^2 \cr
F_{S}& = & U |\Delta_s|^2 -\chi_{1s} \Delta_1 \Delta_s - \chi_{2s} \Delta_2 \Delta_s + \beta_3 |\Delta_s|^4.
\ea
$F_{DOT}$ describes the energetics of the d-wave orbital triplet
pairing, where $\Delta_{1}$ and $\Delta_{2}$ denote the order parameters
of the two d-wave condensates: for example,
$\Delta_{xy} = \Delta_{1}s_{x}s_{y}$ and $\Delta_{x^{2}-y^{2}} = \Delta_{1} (c_{x}-c_{y})$.
The term $\chi_{12}$ describes the attractive Josephson
coupling between the two gap functions
generated by the orbital
Rashba effect, which will tend to phase-lock the two condensates to
produce a fully gapped \spm state. A microscopic calculation  gives
$\chi_{12}= (N (0)/4)\ln (\frac{\omega_{sf}}{2 \pi T})$ (see SOM)
where $N (0)$ is the density of states and $\omega_{sf}$ is the characteristic cut-off
energy scale of the d-wave pairing.

$F_{S}$ describes the effect of the Coulomb
interaction, which imposes an energy cost $U$ associated with any
uniform s-wave order parameter.
The orbital-Rashba coupling
generates a linear coupling between the d- and
s-wave condensates described by $\chi_{1s}$ and $\chi_{2s}$.
For small values of
$\Delta_{s}$, the s-wave term can be integrated out, yielding a
renormalized Josephson coupling
\begin{equation}\label{}
\chi_{12}\rightarrow \chi^{eff}_{12} = \chi_{12 }- \left(\frac{\chi_{1s}\chi_{2s}}{U} \right)
\end{equation}
A microscopic calculation shows that
$\chi_{1 s} = \chi_{2s} \sim  \sum_{I=\pm }
{\rm sgn} (I) ln \left( \tfrac{\omega_{sf}}{2 \pi T} \right)$,
with equal and opposite contributions from the two helical bands. When both bands
cross $E_F$, $\chi_{1s} = \chi_{2s} = 0$,
thereby demonstrating the phase cancelation mechanism; in this case
the Josephson coupling thus
$\chi_{12}^{eff}  > 0$, driving the system into the
energetically favored $s^{\pm}$ state. However, this compensation
fails when the electron band is doped away from $E_F$, and $\chi_{1s}
= \chi_{2s}\sim  - ( N (0)/2) ln \left( \tfrac{\omega_{sf}}{2 \pi T} \right)$. At
this Lifshitz point, the Coulomb repulsion renormalizes the Josephson forcing it to change sign,
and a first-order phase transition from the $s^{\pm}$ to the octet
state will occur. A microscopic calculation gives (see online material)
\begin{equation}\label{}
\chi_{12}^{eff} = N (0)\ln \left(\frac{\omega_{sf}}{2 \pi T_{c}} \right)\times
\left\{\begin{array}{lc}
\frac{1}{4}& (\hbox{electron and hole pockets}),\cr
-\frac{1}{8}& (\hbox{hole pockets only}).
\end{array}
\right.
\end{equation}
We thus expect a quantum phase transition from the low angular
momentum \spm  superconducting state to the high angular
momentum state when the Coulomb repulsion overcomes the internal
Josephson coupling. This is most likely to occur in systems without electron pockets.
KFe$_2$As$_2$ exhibits exactly such a Fermi surface structure, and
experiments on  this material show that it undergoes a first order
transition under pressure from a gapless to a fully
gapped superconductor\cite{TailleferNatPhys2013}.
This transition has been interpreted as a competition between
two fine-tuned d-wave and s-wave pairing mechanisms\cite{TailleferNatPhys2013}.
However, the high to low spin transition of the condensate provides an
alternative way to account for this transition within a single pairing
mechanism.




One of the ways  in which the orbital entanglement of the condensate
can be measured, is using polarized Angle Resolved Photoelectron Spectroscopy (pARPES),
which has been recently used to measure the orbital character of the surface states in the topological insulator, Bi$_2$Se$_3$ \cite{DessauNPhys2013}.
As the orbitally entangled condensate develops,
we predict that orbital anisotropy of the ARPES signal will change
in a very specific fashion.
Polarization dependent  ARPES techniques
determine the momentum-resolved orbital anisotropy of the quasiparticles,
defined by $I_3(\bk) = n_{xz}(\bk) - n_{yz}(\bk)$.
In the superconducting state, Andreev scattering off the
orbitally entangled pairs modifies the orbital
Rashba field.  This is because
Andreev scattering off the orbitally entangled condensate
contains an interband term of
strength proportional to  $\vec{d}_{\bk} \times
\vec{B}_{\bk}$, so that two successive
successive Andreev scattering events give rise to an additional component
to the orbital Rashba field. A detailed calculation of the resulting
orbital anisotropy (see Supplementary Information for details) gives
\be
\label{eq:delta I3}
\delta I_{3}(\bk) = I^{sc}_{3} - I^{n}_{3} \approx \frac{\Delta_{xy}(\bk) \left( \epsilon_{x^2-y^2}(\bk) \Delta_{xy}(\bk) - \epsilon_{xy}(\bk) \Delta_{x^2-y^2}(\bk) \right)}{|\vec{B}_{\bk}|^2 \, \Delta_{sc}(\bk)}
\ee
where $\Delta_{sc}(\bk)$ is the full superconducting gap.
The qualitative angular form of this function
is $\delta I_{3} (\theta )\sim \sin(2 \theta)\sin(4 \theta) \sim$ cos$(6
\theta)$. The overall magnitude is
proportional to $1/\Delta_{sc} (\bk )$, so that the changes in the
orbital character are expected to be greatly enhanced in the octet state.
Fig ~\ref{Fig:spm six fold}  contrasts the
predicted orbital anisotropy for the \spm and octet superconductors.
The vanishing of the gap in the octet state leads to a characteristic
cusp like structure in the orbital anisotropy near the gap nodes,
observation of which would provide a definitive test of our theory.

One of the interesting
aspects of the orbital triplet condensates
involves their internal topology.
The unit $\hat {d}_{\bk}= \vec{d}_{\bk }/|d_{\bk }|$
vector defines a winding number,
\begin{equation}
\label{eq:topological number}
\oint_{\Gamma} \hat{z} \cdot \left( \hat{d}\dg (\bk ) \times \partial_{a} \hat{d} (\bk ) \right) dk_{a} = 2 \pi \nu
\end{equation}
The low-spin $s^{\pm}$\ state and  high-spin octet state
have opposite winding numbers $\nu = +2$ and $\nu = -2$ respectively.
At an interface between these two phases,
the change in topology is
expected to produce gapless Andreev bound-states, loosely
analgous to the Majorana surface states in
$^{3}$He-B
\cite{Volovik2003,VolovikJETP2009a,VolovikJETP2009b}. However, here
spin singlet character of the condensates  will produce a Kramers
doublet of counter-propagating Andreev bound states.
This prediction could be tested using an epitaxially
grown interface between optimally doped and electron or hole-doped samples.

We end by mentioning the effect of including the
additional $xy$ orbitals, neglected in
our initial model of orbital triplet pairing.   To describe the
additional entanglement of these orbitals with the condensate
we must introduce two new orbital isospin operators
$I_4 =
i(\ket{xz} \bra{xy} - \ket{xy} \bra{xz})$ and $I_5 = i(\ket{yz}
\bra{xy} - \ket{xy} \bra{yz})$. Since these operators
involve a change in angular momentum of one unit, they carry internal
angular momentum $I=1$ and to form the $s^{\pm }$ condensate
with net angular momentum $J=L+I=0$
their corresponding form factors must have $L=1$  $p$-wave
symmetry, so that now
\ba
H^{sc} & = & \sum_{\bk} \psi\dg_{\bk, I \sigma} \left( \Delta_{xy} I_1 + \Delta_{x^2-y^2} I_3 + \Delta_{x} I_4 + \Delta_{y} I_5 \right) \gamma_1 \psi_{\bk I \sigma}
\ea
where $\Delta_{x}= \Delta \sin k_{x}$ and $\Delta_{y}= \Delta \sin
k_{y} $.    These extra terms may promote additional
low-to-high spin transitions into gapless  $J=I+L=2$ d-wave states.

In conclusion, we have proposed that the $s^{\pm}$ pairing in
the iron based superconductors derives from an underlying orbital
triplet condensate. Our model allows for the possibility of both
``low''  and ``high'' spin configurations of the orbital triplet pairs
and predicts the development of a distinct  orbital anisotropy
signature in the ARPES spectroscopy in the superconducting phase.
We note that this pairing
mechanism may also be relevant for other multi-orbital superconductors
such as
SrRu$_2$O$_4$.

\newpage

\bibliography{FeAs_Bib_revised}
\bibliographystyle{naturemag}

\begin{itemize}
\item []{\bf Supplementary information } is linked to the online
version of the paper at www.nature.com/nature.

\item[]{\bf Acknowledgements:} We should particularly like to thank
Leni Boscones, Andrey Chubukov, Gabriel Kotliar, Andrew Millis, Peter Orth and Qimiao
Si for discussions related to this work.  We are particularly grateful
to Daniel Dessau for helpful discussions on changes to the orbital
character in the superconducting state.  We acknowledge funding from
DOE grant DE-FG02-99ER45790 (Coleman, Ong),
Deutsche Forschungsgemeinschaft through DFG-SPP~1458                            `Hochtemperatursupraleitung in Eisenpniktiden’
(Schmalian) and grant
NSF 1066293 (Coleman) while at the Aspen Center for Physics.

\item []{\bf Contributions } The authors, T. Tzen Ong, Piers Coleman
and Joerg Schmalian contributed equally in the discussions and
development of the ideas in this paper.
T. Tzen Ong carried out the numerical
calculations of the gap and the orbital anisotropy signal in
ARPES spectroscopy.
All authors contributed towards
the writing of the paper and supplementary materials.

 \item[]{\bf Competing Interests} The authors declare that they have no
competing financial interests.

 \item[]{\bf Correspondence} and requests for materials
should be addressed to Piers Coleman.~(email: coleman@physics.rutgers.edu).

\end{itemize}

\newpage

\noindent {\bf Figure Captions}
\renewcommand{\labelenumi}{{\bf Fig.} \theenumi .}
\begin{enumerate}
\item \label{Fig:Orbital isospin} Orbital helicity and orbital triplet $s^{\pm}$
superconducting state.  Fig.~(a) shows the orbital Rashba vector,
$\vec{n}_{\bk}$ and the orbital character in a simplified two band
model for the iron-based superconductors, in the  extended zone scheme.
The orbital helicity $I=\vec{I}\cdot \vec{n}_{\bk }$ is positive on central hole
pockets (red:$I=+1$) and negative on electron pockets (blue: $I=-1$),
thus the orbital polarization of the hole Fermi surface
is parallel to $\vec{n}_{\bk}$ (Red and blue hollow arrow denotes $xz$ and $yz$ respectively).
Fig.~(b) shows the conventional $s^{\pm}$ superconducting state,
and Fig.~(c) shows the the orbital triplet $s^{\pm}$.  In
conventional $s^{\pm}$, the $\pm$ sign change is determined by the
$cos (k_x) \, cos (k_y)$ form factor in $\bk$-space; in the
orbital triplet state, the sign change depends on the orbital
helicity ($sgn(I)$) of the bands.

\item \label{Fig:Octet SC gap}
{Superconducting gap in ``high"-spin orbital triplet octet
state. (a) octahedral structure of the superconducting
gap on the hole pocket around $\Gamma$, when $\vec{n}_{\bk}$ and
$\vec{d}_{\bk}$ have opposite helicities, (b)
fully gapped electron pocket
around $M$.}

\item \label{Fig:spm six fold}
{Polar plot of orbital anisotropy in $s^{\pm}$ phase and octet phase.
(a) \& (b) show the angular dependence (from $0$ to $\frac{\pi}{2}$) of the
orbital anisotropy $\delta \langle I_3 \rangle$ around the hole pocket
centered at $\Gamma$ for the $s^{\pm}$ state and octet state respectively.
The sin$(2 \theta)$ sin$(4 \theta) \sim$ cos
$(6 \theta)$ structure (dodecagonal) is clearly revealed in both cases,
and the `high"-spin octet state shows a unique cusp-like feature at the gapless nodal points.}

\item \label{Fig:Edge state}{Helical edge states of orbital triplet pairing. Helical gapless Andreev edge states at a domain wall between a low-spin $s^{\pm}$ (left) and high-spin octet (right) state.}

\end{enumerate}
\newpage
\setcounter{figure}{0}

\begin{figure}[t]
\begin{center}
\includegraphics[width=16cm]{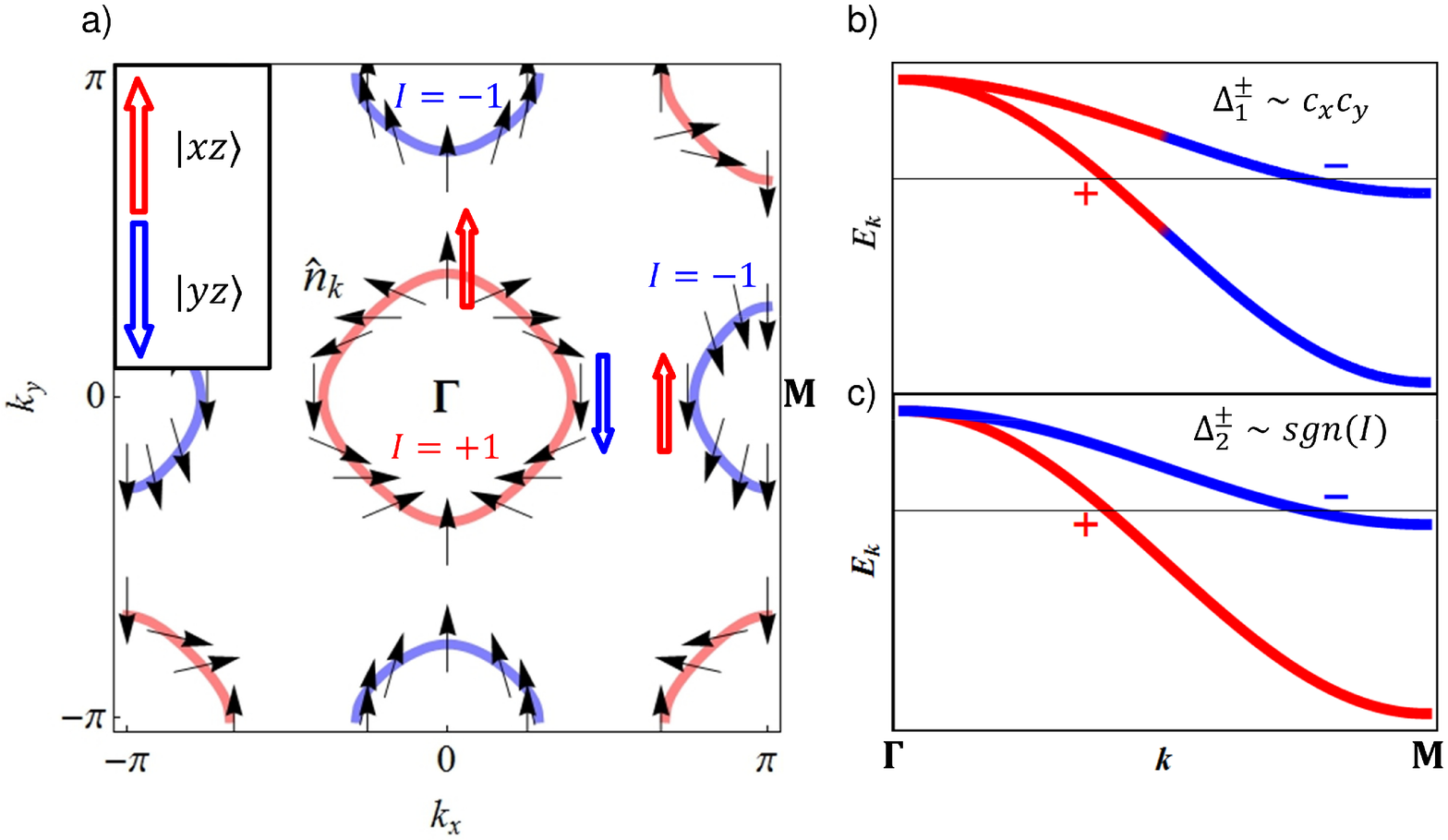}
\end{center}
{\vskip 5cm\Large \sffamily Figure 1.}
\end{figure}
\newpage

\begin{figure}[t]
\begin{center}
\includegraphics[width=16cm]{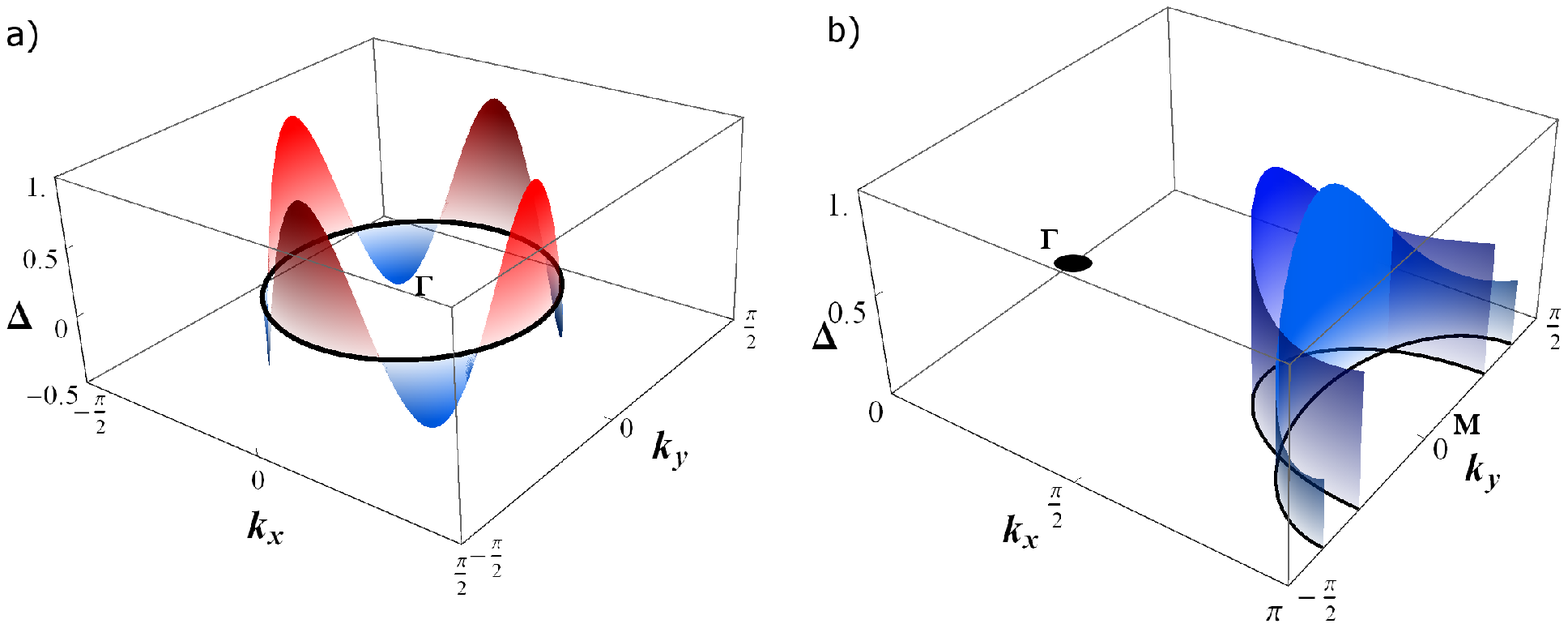}
\end{center}
{\vskip 5cm\Large \sffamily Figure 2.}
\end{figure}
\newpage

\begin{figure}[t]
\begin{center}
\includegraphics[width=\textwidth]{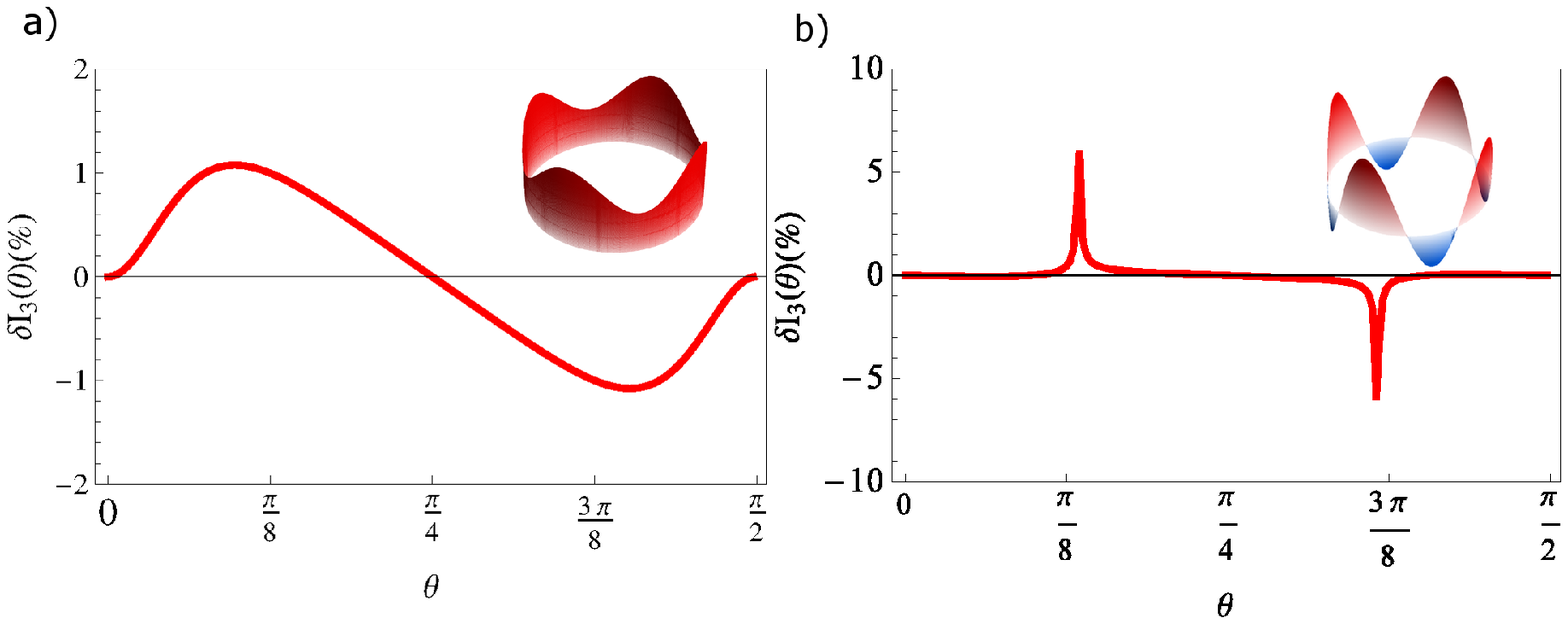}
\end{center}
{\vskip 5cm\Large \sffamily Figure 3.}
\end{figure}
\vskip 4in
\newpage
	
\begin{figure}[t]
\begin{center}
\includegraphics[width=\textwidth]{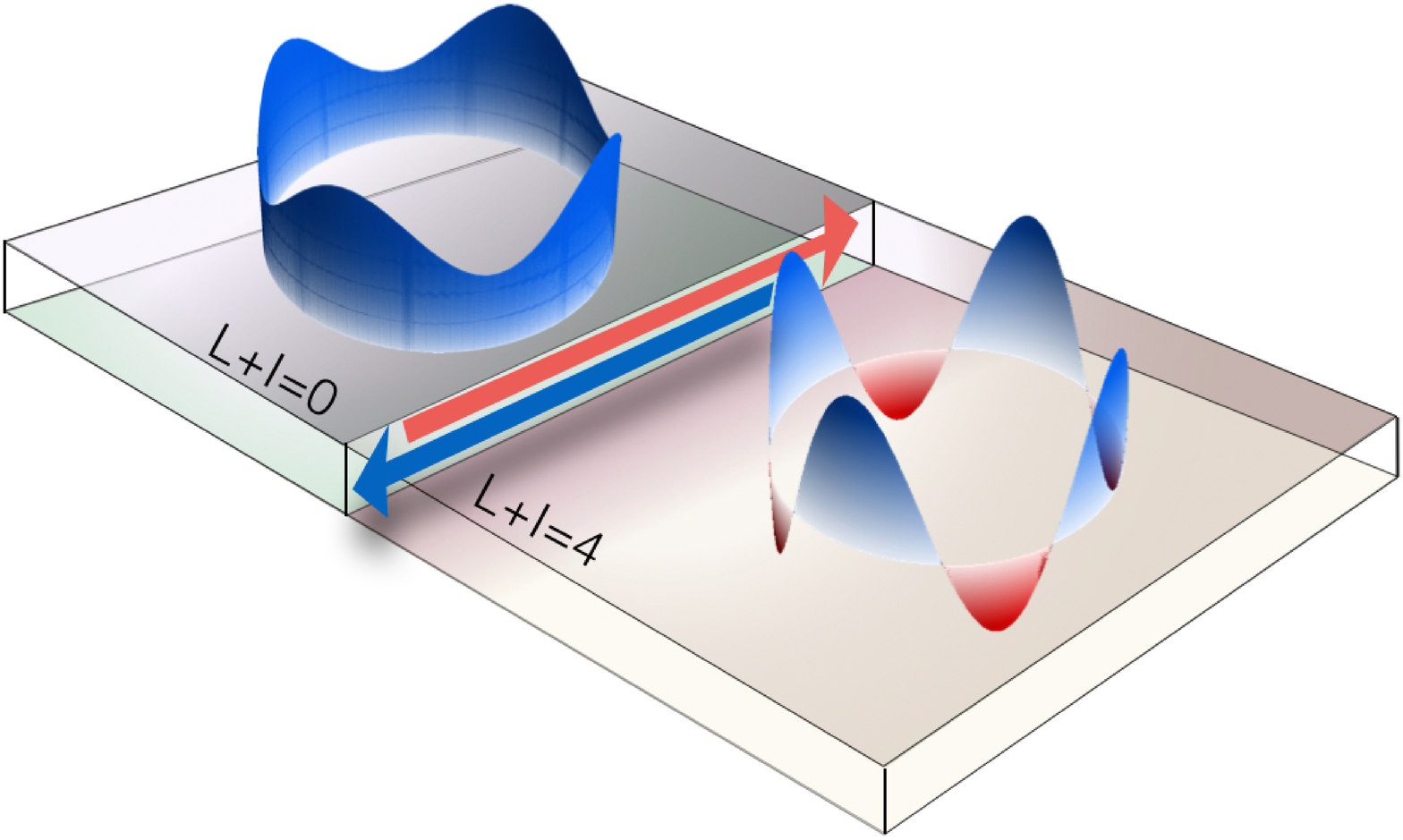}
\end{center}
{\vskip 5cm\Large \sffamily Figure 4.}
\end{figure}

\end{document}


\centerline{\Large\bf
Supplementary Online Material}

\vskip 0.3in
\centerline{\bf
\large Entangled Orbital Triplet Pairs in Iron-Based Superconductors
}
\vskip 0.4in
\centerline{T. Tzen Ong $^{1}$, P. Coleman $^{1,2}$ and J. Schmalian $^{3,4}$}
\centerline{\sl
$^1$Center for Materials Theory,
Rutgers University, 136 Frelinghuysen Rd., Piscataway, NJ 08854-8019, USA}
\centerline{\sl
$^2$Department of Physics, Royal Holloway, University
of London, Egham, Surrey TW20 0EX, UK.} \
\centerline{\sl
$^3$Institute for Theory of Condensed Matter, Karlsruhe Institute of
Technology, 76131 Karlsruhe, Germany.}
\centerline{\sl
$^4$DFG Center for Functional Nanostructures,
KIT, 76128 Karlsruhe, Germany.}

\tableofcontents

\vfill \eject

\section{Landau Free Energy \& Derivation of Renormalized Josephson Coupling}

This section presents  the derivation of the Landau Free energy that describes
the transition between the ``low"-spin $s^{\pm}$ state and the
``high"-spin octet state. We assume an isotropic system in Eq.~(9) of
the main paper to illustrate the key physics of renormalization of the
Josephson coupling $\chi_{12}$ by the Coulomb repulsion; hence the
coefficients for $\Delta_1$ and $\Delta_2$ are equal. In the physical
systems with tetragonal symmetry, the coefficients for $\Delta_1$ and
$\Delta_2$ are generally not equal as they are not related by symmetry
operations of the $C_{4v}$ group. Thus, the free energy would have the
general form,
\ba
\label{app eq: Landau free energy}
F & = & F_{DOT} + F_s, \cr
F_{DOT} & = & \alpha_1 (T-T_{c}) |\Delta_1|^2 + \alpha_2 (T-T_{c}) |\Delta_2|^2 - \chi_{12} \Delta_1 \Delta_2 \cr
        &   & + \beta_1 |\Delta_1|^4 + \beta_2 |\Delta_2|^4 + \beta_{12} |\Delta_1|^2 |\Delta_2|^2 ,\cr
F_{S}& = & U |\Delta_s|^2 - \chi_{1s} \Delta_1 \Delta_s - \chi_{2s} \Delta_2 \Delta_s + \beta_3 |\Delta_s|^4.
\ea
where $\Delta_{1}$ and $\Delta_{2}$ are the gap parameters for the
$d_{xy}$ and $d_{x^{2}-y^{2}}$ orbital triplet condensates.
In the isotropic case described in the main paper, $\alpha_1 =
\alpha_2$ and $\beta_1 = \beta_2$.  The ``low"-spin to ``high"-spin
transition is driven by a change in the relative orientation of
$\vec{d}_{\bk}$ with respect to $\vec{n}_{\bk}$, i.e. by reversing
${\rm sign}(\Delta_{1} \Delta_{2})$. The relative phase of
$\Delta_{1}$ and $\Delta_{2}$, is determined
by the internal Josephson coupling $\chi_{12}$. When both helical
bands cross $E_F$, we shall see that
$\chi_{12} > 0$ while $\chi_{1s}$ and $\chi_{2s}$ are almost zero, so
the ground-state energy is minimized
when $\Delta_{1}$ and $\Delta_{2}$ have the same phase,
forming an isotropic gap with no nodes. By contrast, when one band is
removed from the Fermi surface by doping, then $\chi_{1s}$ and
$\chi_{2s}$ become large, and the effective Josephson coupling becomes
\begin{equation}\label{}
\chi_{12}^{eff}= \chi_{12}- \frac{\chi_{1s}\chi_{2s}}{\chi_{s}} < 0
\end{equation}
causing a reversal of the relative sign of the two condensates and the
formation of nodes in the gap.

We can derive the free energy using a BCS Hamiltonian that includes
the orbital triplet pairing and a repulsive Hubbard $U$ term, which we
factorize into a product of $s$-wave pairing terms using the
Hubbard-Stratonovich method,
\begin{equation}\label{}
H^U = U \sum_{i, I} n_{i I \uparrow}
n_{i, I, \downarrow} \rightarrow \sum_{i, I} \Delta_s c\dg_{i, I,
\uparrow} c\dg_{i, I, \downarrow} + H.C.- |\Delta_{s}|^{2}/U,
\end{equation}
to obtain
\ba
\label{eq:BCS Hamiltonian}
H & = & \sum_{\bk} \psi\dg_{\bk} \left[ \epsilon_s(\bk) \gamma_3 + \vec{B}_{\bk} \cdot \vec{I}
        + \vec{d}_{\bk} \cdot \vec{I} \gamma_1 + \Delta_s \gamma_1 \right] \psi_{\bk}
        + N_s \left( \frac{\Delta_1^2}{g_1} + \frac{\Delta_2^2}{g_2} - \frac{\Delta_s^2}{U} \right),
\ea
where $N_s$ is the number of lattice sites and the Rashba and orbital
d-vectors are defined by
\begin{eqnarray}\label{l}
\vec{ B}_{\bk } &=& (\epsilon_{xy},0,\ \epsilon_{x^{2}-y^{2}}),\cr
\vec{ d}_{\bk }&=& (\Delta_{xy},0,\Delta_{x^{2}-y^{2}})= (
\Delta_{1}d _{xy},0, \ \Delta_{2} d_{x^{2}-y^{2}}),
\end{eqnarray}
where $d_{xy}$ and $d_{x^{2}-y^{2}}$ are d-wave form factors,
for instance $d_{xy}= s_{x}s_{y}$ and $d_{x^{2}-y^{2}}=
(c_{x}-c_{y})$ in a tight binding basis.

 The Bogoliubov spectrum is,
\be
\label{eq:Bogliubov spec}
E^{\alpha }(\bk) = \left[ A(\bk) - \alpha  \sqrt{A^2(\bk) - \gamma^2(\bk)} \right]^{\frac{1}{2}},
\ee
where $\alpha  = \pm 1$ for the two superconducting bands and
\ba
\label{eq:A and gamma}
A(\bk) & = & \epsilon_s(\bk)^2 + |\vec{B}_{\bk}|^2 + |\vec{d}_{\bk}|^2 + \Delta_s^2, \cr
\gamma(\bk)^2 & = & \left[ \epsilon_s(\bk)^2 - |\vec{B}_{\bk}|^2 + |\vec{d}_{\bk}|^2 - \Delta_s^2 \right]^2 \cr
              & & + 4 \left[ \Delta_{xy}(\bk) \epsilon_{xy}(\bk) + \Delta_{x^2-y^2}(\bk) \epsilon_{x^{2}-y^{2}}(\bk) - \Delta_s \epsilon_s(\bk) \right]^2.
\ea

The free energy is then given by,
\ba
\label{app eq:free energy}
F & = & N_s \left[ \frac{\Delta_1^2}{g_1} + \frac{\Delta_2^2}{g_2} - \frac{\Delta_s^2}{U} \right]
      - 2 T \sum_{\bk, {\alpha}} ln \left[ 2\cosh \left(
\frac{ E^{{\alpha}}(\bk)}{2T} \right) \right].
\ea

A Taylor expansion of Eq.~\ref{app eq:free energy} will then give us the coefficients of the Landau Free energy in Eq.~\ref{app eq: Landau free energy}. We will now carry out the calculation around the saddle point, which gives the following coupled gap equations,
\begin{eqnarray}
\label{eq:saddle pt eqns}
\begin{pmatrix}
\frac{1}{g_{1}} - \tilde{\chi}_{11} & - \tilde{\chi}_{12} &  -\tilde{\chi}_{1s} \cr
- \tilde{\chi}_{12} & \frac{1}{g_{2}} - \tilde{\chi}_{22} & -\tilde{\chi}_{2s} \cr
-\tilde{\chi}_{1s} & -\tilde{\chi}_{2s} & -\frac{1}{U} - \tilde{\chi}_{s}
\end{pmatrix}
\begin{pmatrix} \Delta_1 \cr \Delta_{2} \cr \Delta_s \end{pmatrix} = 0.
\end{eqnarray}

Denoting $\int_{\bk} = \int \tfrac{d^2 k}{(2 \pi)^2}$, the Josephson
couplings between the different pairing channels are then given as follows:
\begin{eqnarray}\label{eq:chi1}
\tilde{\chi}_{11} &=& \sum_{{\alpha}=\pm 1 } \int_{\bk} \frac{\rm th(\beta E^{{\alpha} }(\bk)/2)}
{2 E^{{\alpha} }(\bk)} \left[1 - 2{\alpha} \frac{(\epsilon_{x^2-y^2})^{2} + \Delta_s^2
}{
\sqrt{A^2 -\gamma^2} }
\right] d^{2}_{xy},\\
\label{eq:chi2}
\tilde{\chi}_{22} &=& \sum_{{\alpha}=\pm 1 }\int_{\bk } \frac{\rm{th}(\beta E^{\alpha}(\bk)/2)}
{2 E^{{\alpha}}(\bk)} \left[1 - 2{\alpha} \frac{(\epsilon_{xy} )^{2} + \Delta_s^2 }{\sqrt{A^2 -\gamma^2}}
\right] d^{2}_{x^{2}-y^{2}},\\
\label{eq:chis}
\tilde{\chi}_{s} &=& \sum_{{\alpha}=\pm 1 } \int_{\bk} \frac{\rm th(\beta E^{\alpha}(\bk)/2)}
{2 E^{\alpha}(\bk)} \left[1 - 2{\alpha} \frac{(\Delta_{x^2-y^2})^2 + \Delta_{xy}^2 }{\sqrt{A^2 -\gamma^2} }
\right],\\
\label{eq:chi12}
\tilde{\chi}_{12} &=& \sum_{{\alpha}=\pm 1 } \int_{\bk } \frac{\rm{th}(\beta E^{\alpha}(\bk)/2)}
{2 E^{{\alpha} }(\bk)} \left[ 2{\alpha} \frac{\epsilon_{xy} \epsilon_{x^2-y^2}}{\sqrt{A^2 -\gamma^2}}
\right] d_{xy} d_{x^{2}-y^{2}}\\
\label{eq:chi1s}
\tilde{\chi}_{1s} &=& -\sum_{{\alpha}=\pm 1 } \int_{\bk } \frac{\rm{th}(\beta E^{\alpha}(\bk)/2)}
{2 E^{{\alpha} }(\bk)} \left[ 2{\alpha} \frac{\epsilon_{xy}  \epsilon_{s}}{\sqrt{A^2 -\gamma^2}}
\right] d_{xy},\\
\label{eq:chi2s}
\tilde{\chi}_{2s} &=& - \sum_{{\alpha}=\pm 1 } \int_{\bk } \frac{\rm{th}(\beta E^{\alpha}(\bk)/2)}
{2 E^{{\alpha} }(\bk)} \left[ 2{\alpha} \frac{\epsilon_{x^2-y^2} \epsilon_{s} }{\sqrt{A^2 -\gamma^2}}
\right] d_{x^2-y^2},
\end{eqnarray}
where for clarity, we have suppressed the momentum labels.
Evaluating $\tilde{\chi}_{12}$, $\tilde{\chi}_{1s}$ and
$\tilde{\chi}_{2s}$ at $T=T_{c}$ for $\Delta_{1} = \Delta_{2} = \Delta_{s}= 0$ gives
the Landau free energy coefficients $\chi_{12}$, $\chi_{1s}$ and
$\chi_{2s}$ in Eq.~(9) of the main paper
(here we drop the tilde's to denote the evaluation of  these quantities at $T_{c}$). The coefficients
$\beta_{1}$, $\beta_{2}$ and $\beta_{3}$ are given by Taylor
expansions of $\tilde{\chi}_{12}$, $\tilde{\chi}_{1s}$ and
$\tilde{\chi}_{2s}$.
\ba
\label{app eq: Landau coeffs}
\beta_1 & = & -\frac{1}{4} \frac{\partial \tilde{\chi}_{11}}{\partial \Delta_1^2}, \cr
\beta_2 & = & -\frac{1}{4} \frac{\partial \tilde{\chi}_{22}}{\partial \Delta_2^2}, \cr
\beta_{12} & = & -\frac{1}{2} \left( \frac{\partial \tilde{\chi}_{11}}{\partial \Delta_2^2} + \frac{\partial \tilde{\chi}_{12}}{\partial \Delta_1 \Delta_2} \right).
\ea

To demonstrate the renormalization of the Josephson coupling $\chi_{12}$ by the Coulomb repulsion,
we solve for $\Delta_s$ in the third row of Eq.~\ref{eq:saddle pt eqns},
\be
\Delta_s = -
\left(
\frac{\chi_{1s} \Delta_{1s} + \chi_{2s} \Delta_{2s}}{1/U + \chi_{s}}\right).
\ee
Substituting this back into the first two rows, we obtain
\begin{eqnarray}
\label{eq:eff saddle pt eqns}
\begin{pmatrix}
\frac{1}{g_{1}} - {\chi}^{eff}_{11} & - {\chi}^{eff}_{12} \cr
- {\chi}^{eff}_{12} & \frac{1}{g_{2}} - {\chi}^{eff}_{22} \cr
\end{pmatrix}
\begin{pmatrix} \Delta_1 \cr \Delta_{2} \end{pmatrix} = 0.
\end{eqnarray}
where
\begin{eqnarray}
\label{eq:ren chiJ}
\chi^{eff}_{12} = \chi_{12} - \frac{\chi_{1s} \chi_{2s}}{\chi_s}, \quad
\quad  (U \rightarrow \infty),
\end{eqnarray}
is an effective, renormalized $\chi^{eff}_{12}$, while
\begin{eqnarray}
\label{eq:rentwo}
\chi^{eff}_{11} &=& \chi_{11} - \frac{\chi_{1s} \chi_{1s}}{\chi_s}, \cr
\chi^{eff}_{22} &=& \chi_{22} - \frac{\chi_{2s} \chi_{2s}}{\chi_s},
\end{eqnarray}
are the corresponding diagonal susceptibilities.
Since $\chi_{1s}$ and $\chi_{2s}$ have the same sign,
the Coulomb interaction thus {\sl reduces} the
Josephson coupling $\chi_{12}$ between the d-wave condensates
to a smaller value $\chi_{12}^{eff}$.
The phase transition to the octet state occurs when $\chi^{eff}_{12}$ changes
sign.  This will occur when the system is strongly hole or electron
doped, such that only hole or electron pockets survive at the Fermi
surface.

To illustrate the sign change of $\chi_{12}^{eff}$ at
the phase transition, we use using a simplified momentum-independent orbital
Rashba coupling with  $\epsilon_{xy}(\bk) = \tilde{t} \, sin(2
\theta_{\bk})$ and $\epsilon_{x^2-y^2}(\bk) = \tilde{t} \, cos(2
\theta_{\bk})$. In this case, the helical bands are split apart by
$\tilde{t}$ and have a constant density of state $N (0)$.
Note that at $T_{c}$,
\begin{equation}\label{}
E^{{\alpha}} (\bk ) = \left\vert \phantom{\int_{-\infty }^{\infty }}\hskip
-0.3in |\epsilon_{s} (\bk ) |- {\alpha} \ \tilde{t}\ \right\vert =
\left\vert\epsilon_{s} (\bk ) - {\alpha} \ {\rm sgn} (\epsilon_{s} (\bk
))\tilde{t}\right\vert
\equiv \vert \epsilon_{s} - I\tilde{t}| \equiv | \epsilon_{I}|
\end{equation}
where we have introduced the normal-state band index
\begin{equation}\label{}
I= \alpha \ {\rm sgn} (\epsilon_{s}).
\end{equation}
The appearance of the ${\rm sgn}
(\epsilon_{s})$ separating the normal and superconducting state
band indices is important in keeping track of which
susceptibilities have cancelling logarithimic components.  We can then make the
substitutions
\begin{eqnarray}\label{l}
\epsilon_{I} &=& \epsilon_{s} - I \tilde{t}\\
\cr
\frac{\tanh \left(\frac{\beta E^{{\alpha }} (\bk )}{2}
 \right)}{2 E^{{\alpha} }(\bk)}
&=& \frac{\tanh \left(\frac{\beta \epsilon_{ I}}{2}
 \right)}{2 \epsilon_{I}} \\
\frac{2\alpha }{\sqrt{A^{2}-\gamma^{2}}}= \frac{\alpha }{ |\epsilon_{s}| \tilde{t}}
&=& \frac{\alpha \, {\rm sgn} (\epsilon_{s})}{ \epsilon_{s} \tilde{t}} = \frac{I}{
\epsilon_{s}\tilde{t}}= \frac{I}{\tilde{t}
 (\epsilon_{I}+I\tilde{t})
}\\
\frac{2 \alpha \, \epsilon_{s}}{\sqrt{A^{2}-\gamma^{2}}}= \frac{ \alpha \, \epsilon_{s}}{ |\epsilon_{s}| \tilde{t}}
&=& \frac{\alpha \, {\rm sgn} (\epsilon_{s})}{ \tilde{t}} = \frac{I}{
\tilde{t}}
\end{eqnarray}
The s-wave susceptibility is simply
\begin{eqnarray}\label{chis}
\chi_{s}
=\sum_{I} N (0)\int_{-\omega_{sf}}^{\omega_{sf}} d\epsilon_{{I}} \left(\frac{\tanh
\left(\frac{\beta \epsilon_{I}}{2} \right)}{2 \epsilon_{I}}\right)
\approx \sum_{I} N (0) \ln \left(\frac{\omega_{sf}}{2\pi T_{c}} \right).
\end{eqnarray}
When the electron and hole bands are present, $\sum_{I}=2$, but when
we hole dope the system the $I=-1$ term can be dropped, and we write
$\sum_{I}\rightarrow 1$.
The Josephson coupling between the two d-wave condensates is
\begin{eqnarray}\label{chi12}
\chi_{12}
&=&\sum_{I = \pm 1} N (0)\int_{-\omega_{sf}}^{\omega_{sf}} d\epsilon_{{I}} \left(\frac{\tanh
\left(\frac{\beta \epsilon_{I}}{2} \right)}{2 \epsilon_{I}}
\right)\frac{I \tilde{t}^{2}}{ \tilde{t} (\epsilon_{I}+I\tilde{t}
)}\int_{0}^{2\pi} \frac{d\theta }{2 \pi }
(\sin^{2}(2\theta)\cos^2(2\theta))\cr
&=&\sum_{I}\frac{N (0)}{8}\int_{-\omega_{sf}}^{\omega_{sf}} d\epsilon_{{I}} \left(\frac{\tanh
\left(\frac{\beta \epsilon_{I}}{2} \right)}{2 \epsilon_{I}}
\right)\frac{\tilde{t}^{2}}{ \tilde{t} (I \epsilon_{I} + \tilde{t})}
\approx \sum_{I}\frac{N (0)}{8} \ln \left(\frac{\omega_{sf}}{2\pi T_{c}} \right)
\end{eqnarray}
Note that the contributions from the electron and hole bands add together.
Similarly, for the Josephson coupling between the s- and d-wave order
parameters we obtain
\begin{eqnarray}\label{chi1s}
\chi_{1s}
&=&-\sum_{I} \int_{\bk }\left(\frac{\tanh
\left(\frac{\beta \epsilon_{I}}{2} \right)}{2 \epsilon_{I}}
\right)
\left(\frac{I }{ \epsilon_{s}\tilde{t} }  \right)
\epsilon_{xy}\epsilon_{s}d_{xy}
\cr
&=&-
\sum_{I} I {N (0)}\int_{-\omega_{sf}}^{\omega_{sf}} d\epsilon_{{I}} \left(\frac{\tanh
\left(\frac{\beta \epsilon_{I}}{2} \right)}{2 \epsilon_{I}}
\right)
\int \frac{d\theta }{2\pi} \sin^{2} (2\theta )
\cr
&=&-
\sum_{I}\frac{N (0)}{2}I \int_{-\omega_{sf}}^{\omega_{sf}} d\epsilon_{{I}} \left(\frac{\tanh
\left(\frac{\beta \epsilon_{I}}{2} \right)}{2 \epsilon_{I}}
\right)
\approx -
\sum_{I}I\left[
\frac{N (0)}{2} \ln \left(\frac{\omega_{sf}}{2\pi T_{c}} \right) \right]
\end{eqnarray}
and similarly, omitting the intermediate steps,
\begin{equation}\label{chi2s}
\chi_{2s}= -\sum_{I}I\left[
\frac{N (0)}{2} \ln \left(\frac{\omega_{sf}}{2\pi T_{c}} \right) \right].
\end{equation}
so the d-s couplings have {\sl equal and opposite } contributions
from the two helical bands. When both bands cross $E_F$,
$\chi_{1s} = \chi_{2s} = 0$ and thus
$\chi_{12}^{eff} = \chi_{12} > 0$, driving the system into the
energetically favored $s^{\pm}$ state.
However, this compensation
fails when the electron band is doped away from $E_F$, leaving behind
the sole contribution from $I=+1$, so that $\chi_{1s}
=  \chi_{2s} = - (N (0)/2) ln \left( \tfrac{\omega_{sf}}{2 \pi T}
\right)$.  Substituting (\ref{chis}), (\ref{chi12}), (\ref{chi1s}) and
 (\ref{chi2s})
into (\ref{eq:ren chiJ}), we then obtain
\begin{equation}\label{}
\chi_{12}^{eff} = N (0)\ln \left(\frac{\omega_{sf}}{2 \pi T_{c}} \right)\times
\left\{\begin{array}{lc}
\frac{1}{4}& (\hbox{electron and hole pockets}),\cr
-\frac{1}{8}& (\hbox{hole pockets only}).
\end{array}
\right.
\end{equation}
Thus, at the Lifshitz point where the electron pockets disappear,
the Coulomb repulsion causes a
a sign change in $\chi_{12}^{eff}$ and a first-order
phase transition from the $s^{\pm}$ to the octet state will then occur.

\section{Orbital Anisotropy in Superconducting Phase}

The orbital Rashba field mixes the orbital quantum numbers of the quasi-particles in the normal state,
and the $\vec{k}$-space dependence of the rotating $\vec{B}_{\bk}$ vector drives a modulation in the orbital character of the normal state quasi-particles. This shows up as a $d_{xy}$-dependence of the orbital anisotropy $I_3(\bk) = n_{xz}(\bk) - n_{yz}(\bk)$, which has been measured experimentally in polarization-dependent ARPES\cite{FengZhangPRB2011}.

However, the orbital Rashba field is modified by Andreev scattering upon condensation into the superconducting phase as the underlying $d$-wave orbital triplets will have a different orbital entanglement from the normal-state quasi-particles, i.e. $\vec{d}_{\bk} \neq \vec{B}_{\bk}$. Hence, the effective Rashba field in the superconducting phase picks up an additional $\vec{d}_{\bk} \times (\vec{B}_{\bk} \times \vec{d}_{\bk}) \cdot \vec{\alpha}$ component, giving rise to a sin$(2 \theta)$ sin$(4 \theta) \sim$ cos$(6 \theta)$ component in the orbital anisotropy $I_{3}(\bk)$ that can also be measured in polarization-dependent ARPES.

The Hamiltonian of the system is given by,
\ba
\label{app eq: Hamiltonian}
H & = & \sum_{\bk} \psi\dg_{\bk \alpha \sigma} \left( \mathcal{H}^{0}_{\bk} + \mathcal{H}^{sc}_{\bk} \right) ( i \sigma^2 \psi_{\bk \alpha \sigma}) \cr
\mathcal{H}^{0}_{\bk} & = & \left( \epsilon_{\bk} \mI - \vec{B}_{\bk} \cdot \vec{I} \right) \gamma^3 \cr
\mathcal{H}^{sc}_{\bk} & = & \vec{d}_{\bk} \cdot \vec{I} \; \gamma^1
\ea

and the orbital anisotropy can be calculated from the full Green's function of the system,
\ba
\mathcal{G}(\bk, \omega) & = & \frac{1}{\omega - H^{0}_{\bk} + H^{sc}_{\bk}} \cr
 & = & \frac{1}{ \omega - \left( \epsilon_{\bk} \gamma_3 - \vec{B}_{\bk} \cdot \vec{I} \gamma_3 + \vec{d}_{\bk}\cdot \vec{I} \gamma_1 \right) } \cr
 & = & \left( \omega + ( \epsilon_{\bk} - \vec{B}_{\bk} \cdot \vec{I} ) \gamma_3 + \vec{d}_{\bk} \cdot \vec{I} \gamma_1 \right) \cr
 &  & \times \frac{\left( \omega^2 - (\epsilon_{\bk}^2 + |\vec{B}_{\bk}|^2 + |\vec{d}_{\bk}|^2) - 2 \epsilon_{\bk} \vec{B_{\bk}} \cdot \vec{I} + 2 \vec{B}_{\bk} \times \vec{d}_{\bk} \cdot \vec{I} \gamma_2 \right)}{ \left[ \omega^2 - (\epsilon_{\bk}^2 + |\vec{B}_{\bk}|^2 + |\vec{d}_{\bk}|^2) \right]^2 - 4\epsilon_{\bk}^2 |\vec{B}_{\bk}|^2 - 4 |\vec{B}_{\bk} \times \vec{d}_{\bk}|^2 }
\ea
where we first multiply both numerator and denominator by the factor $\left( \omega + ( \epsilon_{\bk} + \vec{B}_{\bk} \cdot \vec{I} ) \gamma_3 + \vec{d}_{\bk} \cdot \vec{I} \gamma_1 \right)$ to trace out the $\gamma_3$ component.
To trace out the orbital matrix $\vec{I}$ components, we next multiply by
$\left( \omega^2 - (\epsilon_{\bk}^2 + |\vec{B}_{\bk}|^2 + |\vec{d}_{\bk}|^2) \right.$
$\left. - 2 \epsilon_{\bk} \vec{B_{\bk}} \cdot \vec{I} + 2 \vec{B}_{\bk} \times \vec{d}_{\bk} \cdot \vec{I} \gamma_2 \right)$. We finally obtain the expression for $\mathcal{G}(\bk, \omega)$, with a denominator that is proportional to identity $\mI$,
\ba
\label{app eq: Greens function}
\mathcal{G}(\bk, \omega) & = & \left( \omega + ( \epsilon_{\bk} - \vec{B}_{\bk} \cdot \vec{I} ) \gamma_3 + \vec{d}_{\bk} \cdot \vec{I} \gamma_1 \right) \cr
 &  & \times \frac{\left( \omega^2 - (\epsilon_{\bk}^2 + |\vec{B}_{\bk}|^2 + |\vec{d}_{\bk}|^2) - 2 \epsilon_{\bk} \vec{B_{\bk}} \cdot \vec{I} + 2 \vec{B}_{\bk} \times \vec{d}_{\bk} \cdot \vec{I} \gamma_2 \right)}{ \left( \omega^2 - (E^{-}_{\bk})^2 \right) \left( \omega^2 - (E^{+}_{\bk})^2 \right) }
\ea
where,
\be
\label{app eq: Bogoliubov bands}
E^{\alpha}(\bk) = \left[ (\epsilon_{\bk}^2 + |\vec{B}_{\bk}|^2 + |\vec{d}_{\bk}|^2) - \alpha \sqrt{ 4 \epsilon_{\bk}^2 |\vec{B}_{\bk}|^2 + 4 |\vec{B}_{\bk} \times \vec{d}_{\bk}|^2} \right]^{1/2}
\ee

Multiplying out the terms in the numerator of $\mathcal{G}(\bk,
\omega)$, we see that there is an additional contribution to the
orbital Rashba field in the superconducting state, $\delta
\vec{B}_{\bk} \propto (\vec{d}_{\bk} \cdot \vec{I} \gamma_1)
(\vec{B_{\bk}} \times \vec{d}_{\bk} \cdot \vec{I} \gamma_2)$ and using
the Fierz identity $\sigma^a \sigma^b = i \epsilon^{abc} \sigma^c$, we
obtain $\delta \vec{B}_{\bk}= \vec{d}_{\bk} \times (\vec{B}_{\bk}
\times \vec{d}_{\bk}) \cdot \vec{I} \gamma_3 $. The effective orbital
Rashba field component in $\mathcal{G}(\bk, \omega)$ is then given by,
\ba
\mathcal{G}(\bk, \omega) & = & \cdots +  \frac{- \vec{B}_{\bk} \cdot \vec{I} \gamma_3 + \vec{d}_{\bk} \times (\vec{B}_{\bk} \times \vec{d}_{\bk}) \cdot \vec{I} \gamma_3 }{ \left( \omega^2 - (E^{-}_{\bk})^2 \right) \left( \omega^2 - (E^{+}_{\bk})^2 \right) }
\ea

We project into the particle (hole) basis by $P_{{p (h)}} = \tfrac{1}{2}(1 \pm \gamma_3)$, and the difference in the orbital occupancy between the $xz$ and $yz$ orbitals is given by $I_3$. Hence, the orbital anisotropy is given by,
\be
\label{app eq: Orbital anistotropy}
I_{3}(\bk) = \frac{1}{\pi} \int_{-D}^0 Im \left[ Tr \left( \mathcal{G}(\bk, \omega - i \delta) \frac{(1 + \gamma_3)}{2} I_3 \right) \right]
\ee
where the integral is done only over the hole pocket around $\Gamma$, hence we choose a lower cut-off of $D = \tfrac{E^{+}(\bk) + E^{-}(\bk)}{2}$. Note that the electron and hole pockets have opposite helicities, and integrating over both pockets will have canceling contributions.

Substituting Eq.~\ref{app eq: Greens function} into Eq.~\ref{app eq: Orbital anistotropy}, the shift in orbital anisotropy due to Andreev scattering is given by,
\ba
\label{app eq: delta I3}
\delta I_3(\bk) & = & \int_{-D}^0 \frac{d \omega}{\pi} Im \left[ \frac{\vec{d}_{\bk} \times (\vec{B}_{\bk} \times \vec{d}_{\bk}) \cdot I_3 }{\left( \omega^2 - (E^{-}_{\bk})^2 \right) \left( \omega^2 - (E^{+}_{\bk})^2 \right) } \right] \cr
   & = & \int_{-D}^0  \frac{d \omega}{\pi} \frac{\vec{d}_{\bk} \times (\vec{B}_{\bk} \times \vec{d}_{\bk}) \cdot I_3 }{\left( (E^{+}_{\bk})^2 - (E^{-}_{\bk})^2 \right) } Im \left[ \frac{1}{\left( \omega^2 - (E^{-}_{\bk})^2 \right)} - \frac{1}{\left( \omega^2 - (E^{+}_{\bk})^2 \right)} \right]
\ea
Since we pick up only the poles in the lower helical band $E^{-}_{\bk}$ due to a lower cut-off of $D$, this gives,
\ba
\delta I_{3}(\bk) & = & \int_{-D}^0 \frac{d \omega}{\pi} \frac{\vec{d}_{\bk} \times (\vec{B}_{\bk} \times \vec{d}_{\bk}) \cdot I_3 }{\left( (E^{+}_{\bk})^2 - (E^{-}_{\bk})^2 \right) } \; Im \left[ \frac{1}{2 E^{-}_{\bk}} \left( \frac{1}{\omega - i \delta - E^{-}_{\bk}} - \frac{1}{\omega - i \delta + E^{-}_{\bk}} \right) \right] \cr
   & \approx & \frac{1}{4 |\epsilon_{\bk}||\vec{B}_{\bk}|} \frac{\vec{d}_{\bk} \times (\vec{B}_{\bk} \times \vec{d}_{\bk}) \cdot I_3}{2 \Delta_{sc}(\bk)}
\ea

Thus, the Andreev shift of the Rashba field gives a signal $\delta I_3(\bk)$ of $O(\tfrac{\Delta}{|\epsilon_{\bk}|})$,
and  we have approximated the superconducting gap in the quasi-particle basis by,
\be
\label{app eq: QP SC gap}
\Delta^2_{sc}(\bk) \approx |\vec{d}_{\bk}|^2 - \tfrac{|\vec{B}_{\bk} \times \vec{d}_{\bk}|^2}{|\epsilon_{\bk}| |\vec{B}_{\bk}|}
\ee
and, we have approximated $E^{\pm}(\bk) \approx \left[ \left( \epsilon_{\bk} \pm |\vec{B}_{\bk}| \right)^2 + \Delta^2_{sc}(\bk) \right]^{1/2} \approx \Delta_{sc}(\bk)$, and $(E^{+}_{\bk})^2 - (E^{-}_{\bk})^2  = 2 \sqrt{ 4 \epsilon_{\bk}^2 |\vec{B}_{\bk}|^2 + 4 |\vec{B}_{\bk} \times \vec{d}_{\bk}|^2}  \approx 4 |\epsilon_{\bk}||\vec{B}_{\bk}|$ near the hole and electron pocket Fermi surfaces.

\begin{figure}
\centering
\begin{subfigure}{0.5\textwidth}
  \centering
  \includegraphics[width=8cm]{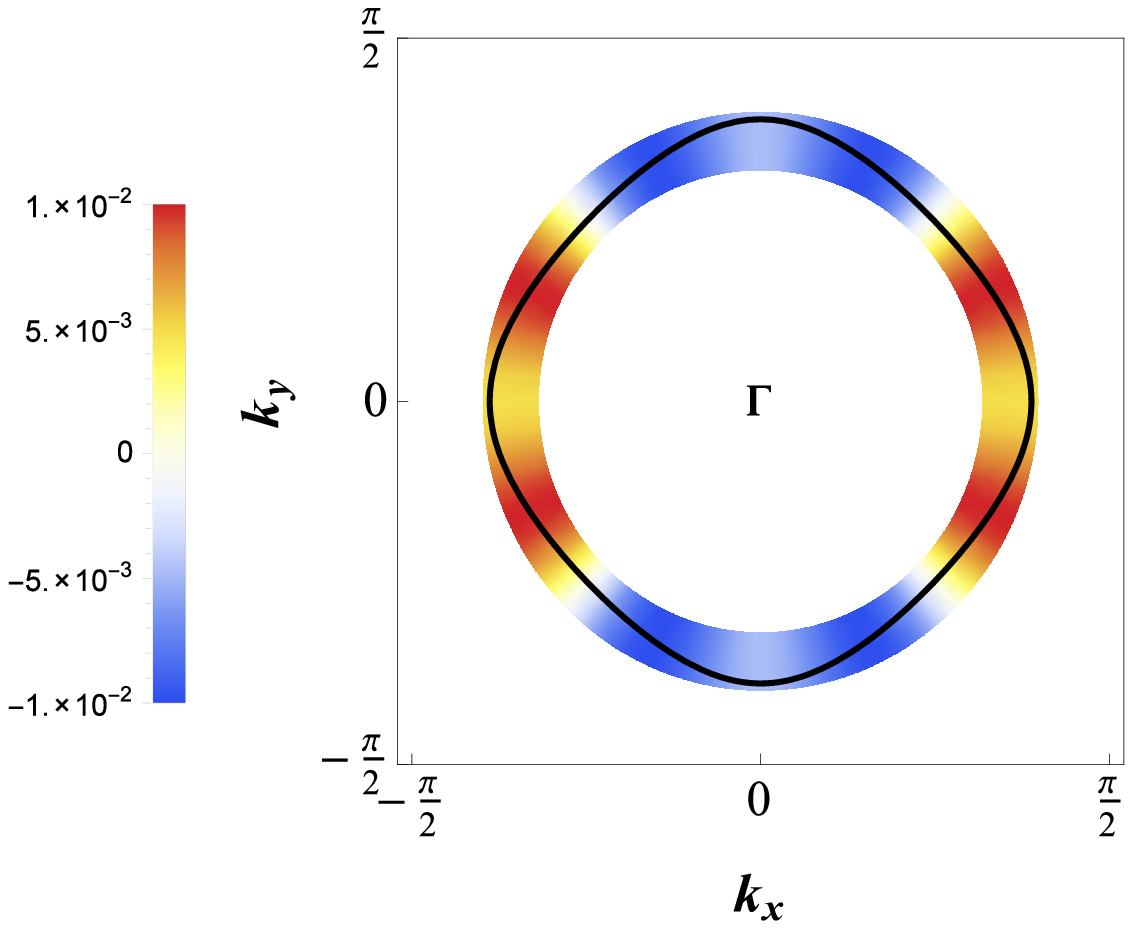}
\end{subfigure}%
\begin{subfigure}{0.5\textwidth}
  \centering
  \includegraphics[width=7cm]{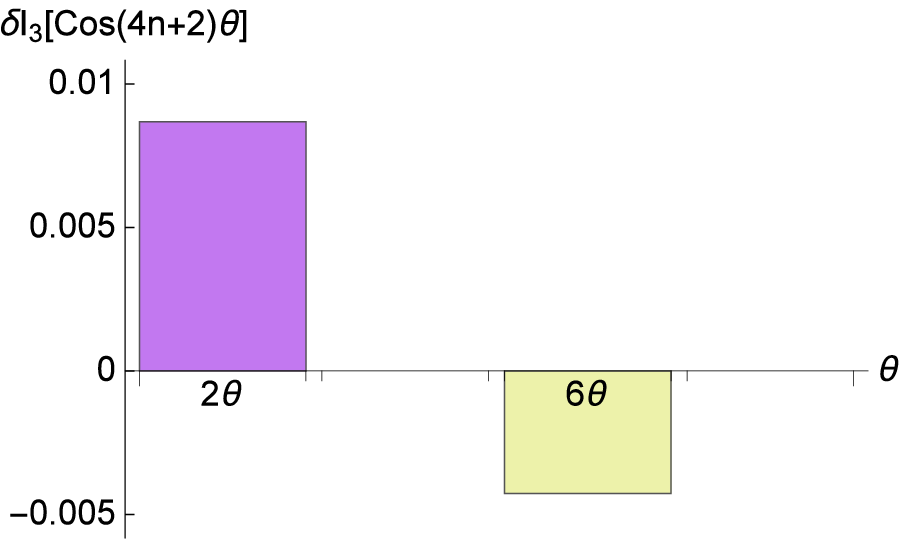}
\end{subfigure}
  \begin{picture}(0,0)
  \setlength\fboxsep{0pt}
  \put(-180,180){\colorbox{white}{\small (a)}}
  \put(0,150){\colorbox{white}{\small (b)}}
  \end{picture}
\caption{Density plot of shift in orbital anisotropy $\delta I_3(\bk) = I^{sc}_3(\bk) - I^{n}_3(\bk)$.
This is the case when $|\Delta_{xy}| > |\Delta_{x^2-y^2}|$,
and the $s^{\pm}$ state shows a negative shift in $I_3(\bk)$, with a
negative cos$(6 \theta)$ component and a positive cos$(2 \theta)$
Fourier component, as shown in Fig.~(b).}
\label{app fig: Delta1 large}
\end{figure}

\begin{figure}
\centering
\begin{subfigure}{0.5\textwidth}
  \centering
  \includegraphics[width=8cm]{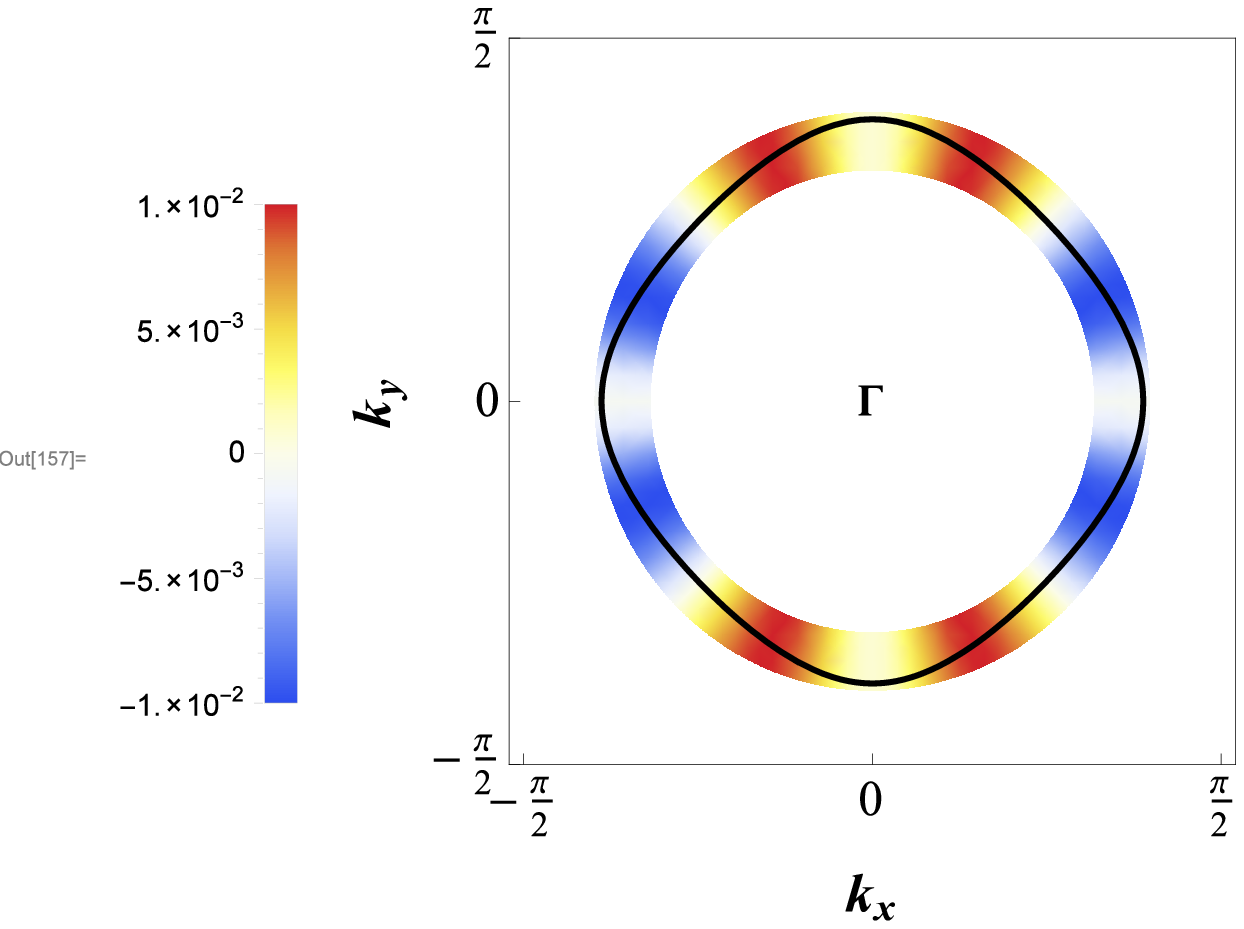}
\end{subfigure}%
\begin{subfigure}{0.5\textwidth}
  \centering
  \includegraphics[width=7cm]{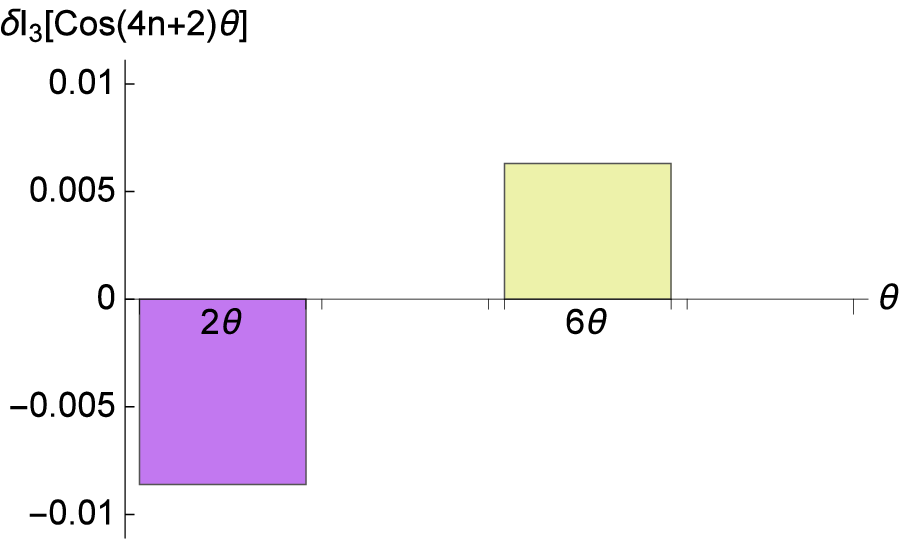}
\end{subfigure}
  \begin{picture}(0,0)
  \setlength\fboxsep{0pt}
  \put(-180,180){\colorbox{white}{\small (a)}}
  \put(0,150){\colorbox{white}{\small (b)}}
  \end{picture}
\caption{Density plot of shift in orbital anisotropy $\delta I_3(\bk) = I^{sc}_3(\bk) - I^{n}_3(\bk)$.
This is the case when $|\Delta_{xy}| < |\Delta_{x^2-y^2}|$,
and the $s^{\pm}$ state shows a positive shift in $I_3(\bk)$, with a
positive cos$(6 \theta)$ component and a positive cos$(2 \theta)$
Fourier component, as shown in Fig.~(b).}
\label{app fig: Delta2 large}
\end{figure}

We numerically evaluate Eq.~\ref{app eq: Orbital anistotropy}, on a $200 \times 200$ grid in the upper right $BZ$ quadrant ${k_x,k_y} \in [{0, \pi/2},{0, \pi/2}]$, and the energy integral over $\omega$ is carried out in Mathemtica using an adaptive algorithim. The orbital anisotropies in the normal and $s^{\pm}$ superconducting phase are plotted in Figs.~\ref{app fig: Delta1 large} \&~\ref{app fig: Delta2 large}, for the two cases of $|\Delta_{xy}| > |\Delta_{x^2-y^2}|$ and $|\Delta_{xy}| < |\Delta_{x^2-y^2}|$ respectively.

We also carry out a Fourier Transform of the orbital anisotropy signal $\delta I_3(\bk)$, and since $I_3 = \langle xz \rangle - \langle yz \rangle$, this means that $I_3 \in B_{1g}$ and it will have nodes along the diagonals, and also has to change sign upon $R^{90^{\circ}}$. Thus, the only Fourier components it will have are $cos 2(2n + 1) \theta$ components, and we show the first two components $cos (2 \theta)$ and $cos (6 \theta)$ which have the largest signals. The $cos (2\theta)$ component will have the largest contribution from the Rashba field $\vec{B}_{bk}$, and the most interesting signal comes from the $cos (6 \theta)$ component that depends on the shift due to the double Andreev scattering process.

\section{Gapless Andreev Edge States}

$C_{4v}$ symmetry guarantees the degeneracy of the $xz$ and $yz$ orbitals,
which allows the system to condense into an orbitally-entangled triplet state.
This non-trivial entanglement is reflected by the winding number of the $\vec{d}_{\bk}$ vector,
which we re-state here for convenience.
\begin{equation}
\label{eq:topological number}
\oint_{\Gamma} \hat{z} \cdot \left( \vec{d}\dg (\bk ) \times \partial_{a} \vec{d} (\bk ) \right) dk_{a} = 2 \pi \nu
\end{equation}

This orbitally entangled nature is reflected in the existence of gapless Andreev edge states
at domain walls between two regions with different topological numbers $\nu$.
Here, we calculate the edge states at the domain wall
between two bulk orbital triplet states of opposite chirality,
with a boundary at $x = 0$ and satisfying the
boundary conditions $\Delta_2(x = -\infty) = -\Delta_2$ and
$\Delta_2(x = \infty) = + \Delta_2$, using the method described by \citet{VolovikJETP2009a}.
The winding number $\nu$ (Eq.~\ref{eq:topological number}) changes sign
from $+ 2$ to $- 2$ across the domain, when $\Delta_2(x)$ changes sign.
However, the orbital Rashba vector $\vec{n}_{\bk}$ remains unchanged
across the domain wall, hence the system is in the ``low"-spin $s^{\pm}$ state
on the left, and in the ``high"-spin octet state on the right of the domain wall.

For small $k_x^2 \ll k_F^2$, we can calculate the edge states perturbatively.
Letting $k_x = k_F + i \partial_x$, $\epsilon_{xy} = \tfrac{t_1}{k_F^2} kx k_y$ and $\epsilon_{x^2-y^2} = \tfrac{t_3}{k_F^2} (k_x^2 - k_y^2)$, we obtain the Hamiltonian,
\ba
\label{eq:Edge Hamiltonian}
H & = & H^{(0)} + H^{'} \\
H^{(0)} & = & i v_F \partial_{x} \gamma_3 + t_3 I_3 \gamma_3 + \Delta_2 (x) I_3 \gamma_1 \\
H^{'} & = & \frac{\Delta_1}{k_F} k_y I_1 \gamma_1 + \frac{t_1}{k_F} k_y I_1 \gamma_3
\ea
where $v_F = \tfrac{k_F}{m}$. $H$ acts on Nambu spinor,
\ba
\psi_{\bk I \sigma} & = & \begin{pmatrix} c_{\bk xz, \uparrow} \cr c_{\bk yz, \uparrow} \cr c\dg_{-\bk yz, \downarrow} \cr - c\dg_{-\bk xz, \uparrow} \end{pmatrix}
\ea
over half the $BZ \in \vec{k} > 0$. Since the Cooper pairs are also spin singlets, there is an additional spin degree of freedom that gives rise to a degenerate Nambu spinor,
\ba
\psi^{'}_{\bk I \sigma} & = & \begin{pmatrix} c_{\bk xz, \downarrow} \cr c_{\bk yz, \downarrow} \cr -c\dg_{-\bk yz, \uparrow} \cr c\dg_{-\bk xz, \downarrow} \end{pmatrix}
\ea

Thus, we can choose to carry out our calculations using only $\psi_{\bk I \sigma}$ over the full $BZ$, which is equivalent to calculations using both $\psi_{\bk I \sigma}$ and $\psi^{'}_{\bk I \sigma}$ over half the $BZ$.

Since $[I_3 \mI, I_3 \gamma_2] = 0$, we can find two zero-energy solutions $H \ket{\psi} = 0$,
\ba
\label{eq:Zero energy modes}
\psi_{\pm}(x)  &=&  {\mathrm exp} \left[ {-\frac{1}{v_F} \int_0^x dx'
\left(\Delta_2(x') I_1 \gamma_1 - i t_3 \alpha_3 \gamma_3 \right)} \right]
\xi_{\pm}, \cr
& & \xi_{+} = \bpm 1 \\0 \\ i \\ 0 \epm  \quad , \quad \xi_{-} = \bpm 0 \\ 1 \\ 0 \\ -i \epm \cr
\ea

Hence, we see that the edge states have a decay length given by $\tfrac{1}{l} = \tfrac{\Delta_2}{v_F}$, and the Fermi momentum along $k_x$ is shifted by $\pm \tfrac{t_3}{v_F}$ for $\xi_{\pm}$ respectively due to the Rashba coupling.
We now treat the edge Hamiltonian $H^{'}$ as a perturbation, and acting in the subspace of $\xi_{\pm}$, this gives

It is straightforward to show that the zero-energy modes satisfy the following Hamiltonian along the edge, and disperse linearly.
\be
\label{eq:Dirac Hamiltonian}
\begin{bmatrix}
H^{'}_{++} & H^{'}_{+-} \\
H^{'}_{-+} & H^{'}_{--} \\
\end{bmatrix}
=
\begin{bmatrix}
0 & v k_y + i \delta k_y \\
v k_y  - i \delta k_y & 0 \\
\end{bmatrix}
\ee
where,
\be
\label{eq:edge velocity}
v = \frac{t_1}{k_F} \quad , \quad \delta = \frac{\Delta_1}{k_F}
\ee

Solving the edge Hamiltonian, Eq.~\ref{eq:Edge Hamiltonian}, gives the following two fermionic zero modes,
\ba
\label{eq:Majorana fermions}
H^{'} \psi_{L,R} & = & \pm \; c k_y \psi_{L,R} \cr
c & = & \sqrt{v^2 + \delta^2}
\ea
where $\psi_{1,2}$ are two linearly dispersing gapless Andreev bound states.
As $\psi_{\bk I \sigma}$ and $\psi^{'}_{\bk I \sigma}$ are spin-polarized states with spin up and down respectively,
we get two sets of gapless Andreev bound states with definite spin and orbital iso-spin,
but these Andreev edge states are not Majorana fermions as they are spin-polarized.

\begin{figure}
\begin{center}
\includegraphics[width=8cm]{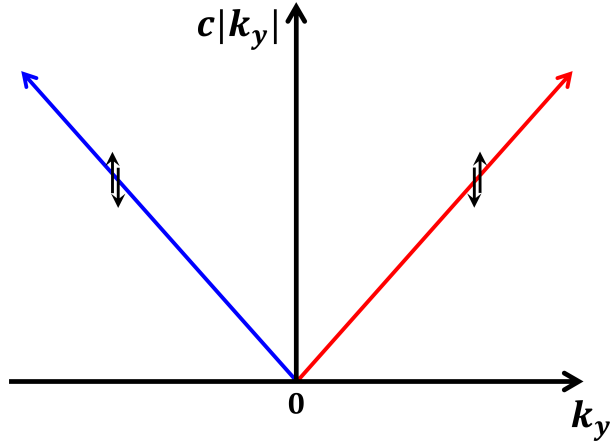}
\end{center}
\caption{Gapless Andreev Edge States.
Linearly dispersing gapless Andreev edge states occur at the domain wall between a ``low"-spin $s^{\pm}$ state and a ``high"-spin octet state. These edge states carry a definite angular momentum of $\langle L_z \rangle = \pm 1$ for the left and right movers. Due to the spin singlet nature of the Cooper pairs, these edge states are also doubly degenerate in spin-space, with a pair of spin-polarized left movers, and another pair of spin-polarized right-movers.}
\label{app fig: Edge state dispersion}
\end{figure}

\bibliography{FeAs_Bib_revised}